# Enhancing Wireless Sensor Network Security through Integration with the ServiceNow Cloud Platform

Syed Atif Ali
Cisco CCIE
Taxes, USA.
Syed.ali@technologyyours.com

Salwa Din
York University, ON
Toronto, Canada.
Email: salkammd@my.yorku.ca

## Abstract

Wireless Sensor Networks (WSNs) continue to experience rapid developments and integration into modern-day applications. Overall, WSNs collect and process relevant data through sensors or nodes and communicate with different networks for superior information management. Nevertheless, a primary concern relative to WSNs is security. Considering the high constraints on throughput, battery, processing power, and memory, typical security procedures present limitations for application in WSNs. This research focuses on the integration of WSNs with the cloud platform, specifically to address these security risks. The cloud platform also adopts a security-driven approach and has attracted many applications across various sectors globally. This research specifically explores how cloud computing could be exploited to impede Denial of Service attacks from endangering WSNs.

WSNs are now deployed in various low-powered applications, including disaster management, homeland security, battlefield surveillance, agriculture, and the healthcare industry. WSNs are distinguished from traditional networks by the numerous wireless connected sensors being deployed to conduct an assigned task. In testing scenarios, the size of WSNs ranges from a few to several thousand. The overarching requirements of WSNs include rapid processing of collected data, low-cost installation and maintenance, and low latency in network operations. Given that a substantial amount of WSN applications are used in high-risk and volatile environments, they must effectively address security concerns. This includes the secure movement, storage, and communication of data through networks, an environment in which WSNs are notably vulnerable. The limitations of WSNs have meant that they are predominantly used in unsecured applications despite positive advancements. This study explores methods for integrating the WSN with the cloud.

# 1. Introduction

Wireless Sensor Networks (WSNs) are increasingly incorporated into a wide range of technological applications. Despite their utility, WSNs are vulnerable, making them a target for attacks. Early detection is critical given the distributed nature of these networks, but many vulnerabilities do not have readily available solutions [1]. The consequences for systems that provide critical services, such as utilities, transport, and communication systems, can be severe. Reduced availability, integrity, or confidentiality might severely impact victims' operations, revenue, and stakeholders' confidence. Furthermore, the frequent requirements for cost-effectiveness and power-constrained operation pragmatically impact the implementation of both security solutions and host-based remediation efforts to detect, isolate, and recover from such compromises. The result is a call for a security regime that complements the traditional network and host approaches [2][3][4]. This context motivates this research, which considers the security of WSNs from the perspective of their interactions with proposed technological solutions drawn from multiple areas. A number of solutions seek to blend WSNs with cloud solutions. Given that these solutions currently provide interface functionality such as storage, event locations, or object query repositories for WSNs, attention is turning to issues of cloud-led inter-device coordination. We postulate that for these WSN-cloud hybrids, securing the cloud component using known or emerging cloud security standards can assist in certain security tasks for WSNs. We are currently analyzing standards as a solution to enhance the security of WSNs in cloud and WSN technology couplings [5].

## 1.1. Background and Motivation

Wireless Sensor Networks (WSNs) have a long history. Often, a WSN consists of nodes—each in part a sensor and part of an ad hoc multi-hop wireless network—deployed in an area of interest, collecting data. A WSN captures procedures and relations followed by fields as diverse as civil, electrical, environmental, and mechanical engineering, as well as various aspects of computer science [6]. Over the past decade, designs, movable RF antennas, DSP chips, and MEMS technologies have found economic applications. Wireless sensor networks have proven inexpensive and useful in reaching goals. Data reliability is claimed to be reasonable within these fields. These days, considerable research attention is given to WSN technologies.

More scholarly interest is gained by introducing certain constraints with security vulnerabilities. Strong concerns on WSNs inspire the need to enhance their weak native security with innovative technology. The data aggregator in a multi-hop route requires complementary countermeasures to existing WSN security. This aspect is not a major emphasis in the study. Integration of WSNs with ServiceNow: In this work, we propose to build WSN and ServiceNow systems that enhance the synergy of technology [7][8]. This work implements a concept to integrate WSNs through a management station with the analytical tool. This method of exposure helps develop an idea. The basic principle in the study is based on cloud platforms that share a system in a network with the management hosting the WSN to implement the growing need for the current generation. In other words, WSN analytics are passed to ServiceNow as an additional layer to manage security in firm, distributed control systems. Additionally, from the IoT domain, WSN and cloud platforms are in line with a gap in the survey. The focus is based on theory, and the investigation of experimental contrast should provide some support for the challenge [9].

### 1.2. Research Objectives

Through the topic, several research objectives were defined and targeted to reach the intended aim and goal. The study investigated the security challenges in Wireless Sensor Network (WSN) systems and the useful integrated Information Technology Service and ITSM Systems for securing WSN systems. Moreover, the study aims to contribute to enriching the body of knowledge within the field of WSNs and cloud platforms by proposing a methodical solution to enhance WSN security through integration into a cloud platform. Consequently, the contribution is intended to benefit practitioners, showing them the security aspects facing recent trends in IoT (WSNs), which include edge and fog computing using artificial intelligence. Hence, the integration of a WSN with a cloud platform completes another aspect of securing the WSN system by securing the management and service desk part. Measurable results: research objectives must be defined in a smart way, and thus, the contribution must be measurable. Therefore, the objectives of the study are set to measure the effectiveness of the cloud platform. Thus, key performance indicators must be selected to reflect the effectiveness of the platform. In the study, the principle of IoT generations and how they ensure the security and challenges of the fourth generation in terms of the smart learning agent is discussed. Finally, the integration between the WSN and a cloud service desk is introduced to enhance the security of wireless sensor networks, especially the fourth generation and beyond. Furthermore, the paper aims to utilize a common, easy security service desk to secure wireless networks.

## 2. Wireless Sensor Networks

Wireless Sensor Networks (WSNs) have advanced from being suitable for a certain type of application to a technology that is versatile for use in smart cities, military surveillance, environmental impact assessment, precision agriculture, home automation, and healthcare systems. The sensor nodes in WSN can execute their tasks and are capable of transmitting and receiving control signals from the given environment using sensors. WSN provides several advantages when compared to existing technologies for data transmission and processing due to advancements in sensor technology, boosting its reliability and endurance. WSNs are developed using a set of contiguous platforms or sensor nodes with functions ranging from data transmission, storage, processing, to decision-making abilities. WSN typically includes three entities: namely, sensing unit or sensor nodes, base stations, and management [10]. The management component is for managing the entire WSN with programs stacked into the nodes as well as for integrating protocols and operating systems. Sensor nodes can be programmed and have a coaxial wire transceiver that can transmit data to every node. The base station collects and accumulates data from multiple nodes using RF WSNs. The major objective of these WSNs is to establish cost-efficient monitoring and control units and to distinguish unusual activities in an environment. Nonetheless, WSN poses some major concerns, especially when confronted with the open environment such as the internet, which may considerably affect their consistency as well as their data integrity. Such conviction problems become far more acute in some privacy-critical environment applications, such as healthcare systems [11]. Therefore, in a WSN scheme, a pervasive attack can paralyze the complete network and fix its actions, making them hazardous to the related application logic.

## 2.1. Overview of WSNs

Wireless Sensor Networks (WSNs) are an emergent and important technology in cyberspace. There are five main properties of WSNs; distributed properties are the most critical, with these properties making WSNs unique and essential. These are the distributed, scalable, adaptable, versatile, and robust components. The sensor nodes measure and collect physical and environmental data, known as the multimedia signal, from real-time systems. The sensor nodes in a sensor network are usually energy-constrained, which limits their processing capability. The collected physical data are communicated with other sensor nodes. The nodes are usually responsible for creating sub-optimal data paths to move data through the network. The data collected by sensor nodes is passed on to a base station called the data sink.

The nodes of the sensor networks are easy to contaminate and hence may deteriorate due to adverse conditions such as compromise, loss of power, collision, and other causes. Moreover, the entities do not have a fixed network topology, a strong central processing unit, or massive hard disk, which reduces the computational power and storage of the WSN component nodes. However, these are not limitations and can consist of using local processing and data processing facilities to take advantage of parallelism [12]. The typical WSN has two basic components: sensor nodes and a base station. Sensor nodes are the distributed sensors and data collector devices that constitute the network resource components. Sensor nodes also consist of three major components: physical sensing, communication protocols, and data collection and processing elements. WSNs can perform their functions in a variety of settings, such as on Earth's surface, in water, and below the ground. Nowadays, due to artificial intelligence and machine learning, the usage of WSNs is attracting more concerns in the fields of healthcare, intelligent transportation, weather forecasting, intrusion detection, fault diagnosis systems, oceanographic and military applications, and environmental monitoring in disaster management centers, instrumentation, and control systems [13]. As a result, activity in WSNs has grown from research and education to industry, especially using various sensors and cloud technology. Research on WSN technologies has not only focused on critical aspects such as the collection and aggregation of sensory data and energy efficiency but also on addressing numerous security and privacy challenges and the integration of WSN technologies with the cloud. The complexity of WSN systems is determined by their architecture and functionality.

## 2.2. Security Challenges in WSNs

2.2.1. Data Security One of the most significant issues in WSN security is that data can be intercepted and maliciously used to decrease the security of a network. Practical methods that adversaries may use include changing the routing information, energy level of a node, and forcing the distribution to be much skewed [14]. This could ultimately cause the network to enter a halting state.

2.2.2. Unauthorized Access Considering the power constraints of the sensor device, identification and authentication, low power, and secure device communication are considerable challenges. Unauthorized access by gaining full control of sensor motes with high energy, which might be used for hard attacks, internal or external invasions to steal, disrupt, or misuse sensory data, is a concern [15].

2.2.3. Denial of Service (DoS) In a WSN, as all sensor devices are battery-limited nodes, the resources available in every sensor device are very rare. In order to limit the high usage of these resources, many cryptographic functions and algorithms are used to minimize the flood of unnecessary data on the network as well as to make other external communication secure between the nodes. In the past, many network security systems were designed and implemented to secure the network and communication between the nodes. In the case of security vulnerabilities, social acceptance may be purchased at the cost of data integrity in any compromise [16]. From the aspect of different service scenarios, a number of key technologies are addressed in this context, and the technical limitations of related techniques are introduced. The Internet of Things has been proposed as a new paradigm in the field of wireless sensor networks. However, WSN faces many issues due to a lack of physical security.

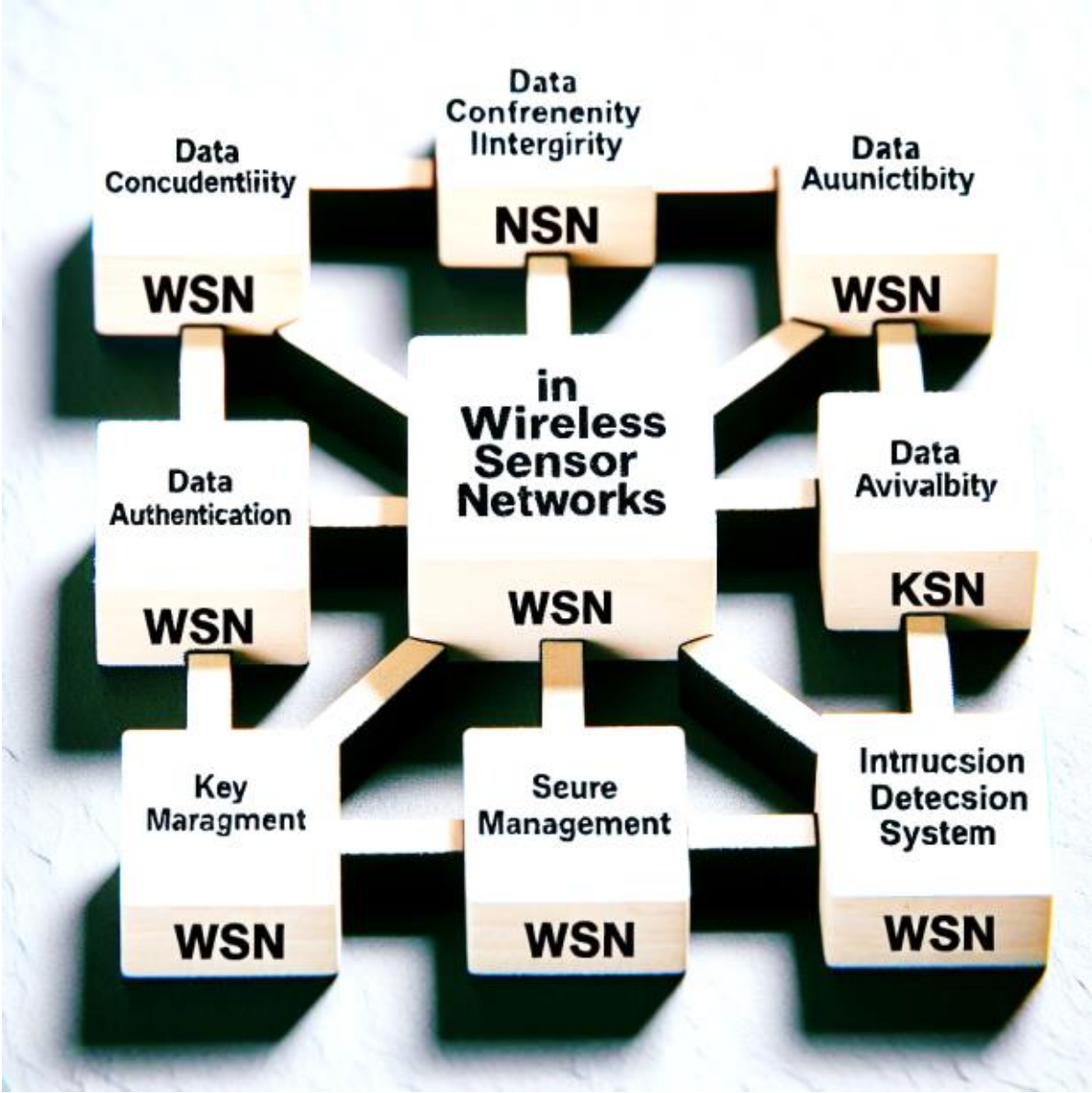

Fig. 1. Security Challenges in WSN.

## 3. ServiceNow Cloud Platform

ServiceNow is a cloud-based service management platform that provides information technology, human resources, and other business support services. Its service model focuses on the management of the service lifecycle, automating the workflow to generate increased efficiency, enterprise visibility, and cost reductions. Furthermore, ServiceNow includes modules enabling the linkage and automation between the service management processes to decrease the lead times for both incidents and changes [17]. This significantly increases the opportunities to detect malicious actions, whether accidental or intentional. It is possible to adapt ServiceNow to various operations. The platform includes a graphical workflow tool allowing customers to automate their processes [18]. Apart from the out-of-the-box solutions, there are many integration points with various solution providers. The solution also includes service reporting featuring a quality and performance control model so that efforts for security effectiveness can be monitored. This solution drives higher business profitability.

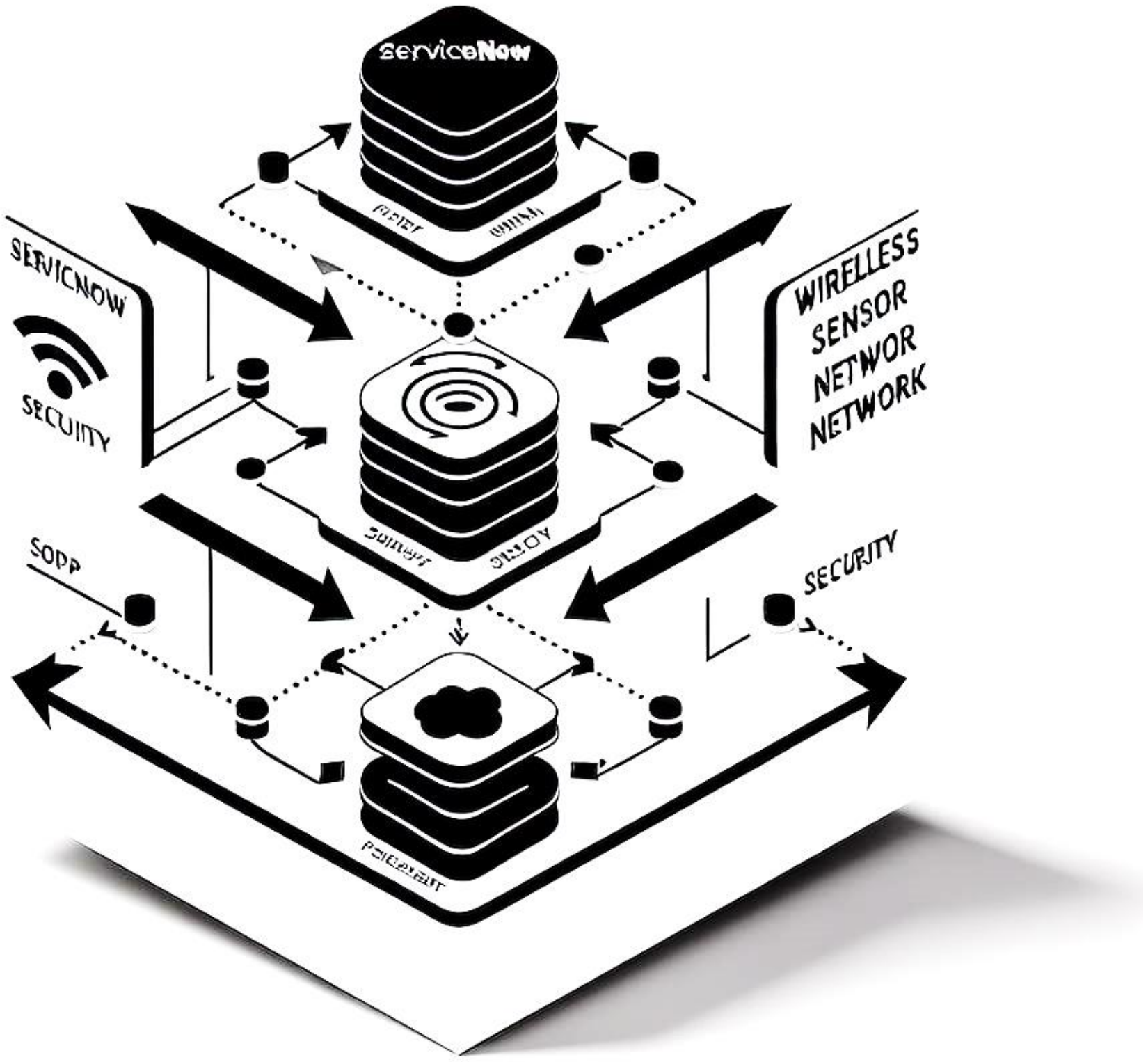

Fig. 2. Conceptual Architecture of WSN with ServiceNow

In the case of a WSN, there is an enormous amount of data at our disposal. Within this big data, there is considerable value. By combining software and know-how for data management, it is possible to transform data into valuable information. This information will be processed by the Decision ServiceNow and potentially made available for business decisions. This is achieved through a combination of different kinds of ITIL service operations. To make the transition from ServiceNow to WSN operation and management as smooth as possible, implementation is done in steps to provide the opportunity to validate configurations, including the network e-social platform [19]. If, for example, an event message is discovered that is suspected of being malicious, an incident and service request is sent automatically through the system to a security administrator who can be a privileged member. A report on the event could be generated and sent to the risk management administrator or another entry point of the WSN computerized analysis. By interfacing a WSN to ServiceNow, several challenges may be overcome. ServiceNow is empowered to provide a solution for a number of security-related tasks challenging WSN administrators. For example, with many monitoring events in the event flow, statistically safe sampling can be a central practice to specify WSN safety health KPIs, including risk register risk by service asset. With the ServiceNow Sensor Solution, the WSN sensor array data provides accessibility and best practices for service management [20].

### 3.1. Overview and Capabilities

Overview and Capabilities. The ServiceNow Cloud Platform is built on a multi-instance architecture that supports strong data isolation. The ServiceNow platform contains multiple components that offer core functionalities. The entire platform is based on the concept of a central configuration management database (CMDB), which contains the structural representation of entities in the world ServiceNow is concerned with. The CMDB is then populated with real-world operational records from data sources that are both part of the platform and can be integrated with external systems.

ServiceNow contains facilities for workflow automation that permit the management of the lifecycle of operational records from any source, with audit and traceability. It includes user interfaces for operational records of any type, allowing non-technical users such as end-customers or fulfillment staff to interact with the system via web browsers or mobile devices. It includes data management facilities for providing real-time insight into data residing on the platform. Thus, ServiceNow can be viewed as a decision support platform as well as an operational platform. ServiceNow is designed for flexibility and can be configured to meet requirements for different industries in order to provide real-time information that can drive appropriate service operations. The architecture of the ServiceNow platform is designed to be scalable and only requires a modern web browser to be leveraged across a range of operations. ServiceNow accommodates Service-Oriented Architectures by providing web services that can be used in integration scenarios at the discretion of an organization.

### 3.2. Integration with Other Systems

ServiceNow is a complete cloud platform that provides scoped business capabilities centered around a rich service architecture. Designed with web services, it offers a powerful integration framework for different kinds of technologies including Java, databases, LDAP, email servers, and business process management systems. Wireless Sensor Networks (WSNs) are a class of wireless networks that allow the integration of sensors that might gather information related to the environment, industrial processes, smart cities, and domotics, with the possibility of further use of the WSN data to generate specific alerts about several applications and to provide an ample range of related services. WSNs have become very popular and required in different applications, playing a determinant role, whether in industrial applications or other domains. Given the importance of WSNs for a variety of applications, there is an urgent need for enhancing data security, uniformity of data management, and sensor data interpretation, as well as operational data acquisition in a variety of domains, especially those related to the Industrial Internet of Things and the Internet of Things. This will facilitate the interoperability of ServiceNow with different industrial sectors and even with other enterprises automatically based on ServiceNow capabilities provided in different services, and also improve the flow of many WSN domain applications.

There are different methods and protocols that make the integration between WSNs and ServiceNow possible in real applications. The presented study aims to highlight the data fusion of WSNs as an extra feature and capability of ServiceNow to integrate with and enhance several domains. The importance of the integration and fusion of the WSNs with ServiceNow can be summarized as follows: Enhance several domains and improve their workflow. Generate quick reports and dashboards for both technical and non-technical people in the companies or the end services. Enhanced data analysis due to fusing the WSN signals and providing easy tools for data analysis and decision-making. Manage alarms from WSN sensors using ServiceNow workflow and management. Automate system management to provide periodic co-management of the reports and dashboards. Although the integration of sensors into ServiceNow can provide a large number of benefits, such migration also raises integration challenges. It is crucial to look ahead to avoid any conflicting issues and time redundancy during the integration of ServiceNow in a timely manner. These features will allow the easy and full integration of WSNs and ServiceNow, starting from the first, opposing the classical point of view with ICT moving from one to another after installation for easy ICT integration with low investment.

## 4. Integration of WSNs with ServiceNow

WSNs can be connected with cloud platforms, which are beneficial in terms of network security by automating different functionalities. The integration of Smart Environment WSNs with the cloud platform is used for data flow, data processing, and decision-making based on the data, e.g., incident identification, qualification, and definition. Due to Smart Environment WSN applications, which are mostly critical in terms of security and the decision-making process, WSNs send data to the cloud platform for real-time monitoring.

WSNs can be integrated with the cloud platform using a specific software platform that can be used to support business process applications. This platform provides form-based data input, database, and application workflow capabilities. To integrate

WSNs with the cloud platform, a specific type of platform is required, which is integrated with the cloud platform to set integration parameters. Therefore, we propose integrating WSNs with the cloud platform. Integration is used to send real-time WSN data to the cloud platform using the File Import Set Table. This section discusses the practical Security-as-a-Service use case, showing the significance of using the cloud platform in a Smart Environment, because the field of Smart Environment consists of WSNs. Since the environment consists of WSNs, which contain a huge amount of data, the first priority of any smart environment is the real-time monitoring of data and automation. Because of monitoring in real-time, if any unusual data is found in the network, it can be handled by the cloud platform. Using the cloud platform, it becomes very easy to send end devices' real-time data to the NOC of the cloud platform, which is designed according to the requirements of the Smart Environment application. Therefore, by integrating with the cloud platform, a WSN can send its data to the cloud platform and perform the required operations in objects, the Incident Management process, the Events process, and the Import Sets process. With this integration, a Smart Environment gains greater visibility. It has been observed that approximately 25% of cloud data is embedded in the platform, providing a good All-in-One kind of solution. With the integration of WSN into the cloud platform, the following benefits have been observed: 1) Improved incident response 2) Improved IoT devices' monitoring 3) Data accuracy and correctness for event generation 4) Data management capability for configuration items and IoT devices 5) Cost and time-effectiveness.

However, in the practical integration of the given case study, we may see some challenges. First, if we use protocols or agreements for integration that are not compatible with each other, the data within the cloud platform will not remain consistent. For example, WSNs send their data to the cloud platform. These fields with default names are automatically mapped in the cloud platform for the import set process schema table. If the field names in the WSN application are different, a custom transformation script would be needed. Also, if WSN data features are constantly changing, then the cloud platform needs to be synced with new features that are being imported. The difficulty of handling a large amount of data is the next challenge in practical integration. Through WSN devices, we can get a large amount of data from the real world, but this data needs to be processed, filtered, and stored in the cloud platform. There are also data management difficulties, such as data handling, data objects, and relationships across all fields.

## 4.1. Benefits and Advantages

Integrating Wireless Sensor Networks (WSNs) with the ServiceNow Cloud Platform can revolutionize wireless network security. In general, wireless sensor network security threats greatly affect the quality of service provided by the network to the users in many aspects. Hardware interfaced between the wireless sensors and network infrastructure functions provides a high level of security. Some of the key advantages of integrating WSNs with ServiceNow are discussed as follows: Security is one of the main benefits of integrating WSNs with ServiceNow. A large number of threats and vulnerabilities can be detected in near real-time with this integration as ServiceNow maintains WSN in the form of CMDB. An automated query raises an alert to the WSN operator upon detecting unexpected traffic in association with recent access added to the WSN infrastructure.

At the same time, this level of integration dramatically simplifies network security management. An integrated WSN can help easily discover potential attacks, for example, worms, and can quickly install the necessary patches through the same interface used for threat detection without the intervention of a network administrator. With such technologies, the WSN security services will not confirm or deny the host's response but will recommend response actions based on the WSN environment. The integration between WSN and ServiceNow also facilitates operational continuity, cost containment, compliance reporting, and ultimately business success. The benefits are summarized as follows: A number of advantages arise from the integration between WSN and ServiceNow. The advantages come in three primary areas: improvements in security and thereby data assurance; improvements in system performance and reliability due to having a better understanding of the network environment where WSN is operating; and enhanced cooperation and data sharing for timely detection and response in the event of a network security incident.

## 4.2. Challenges and Limitations

As the integration is based upon a WSN operating on IoT-based equipment and service assets, there could be many practical and real-time challenges. One such issue is the compatibility of the WSN and IoT technology with the latest Cloud Platform. Also, considering the data privacy infrastructure in Western countries, the WSN and integration may pose privacy problems. The number of WSN devices transmitting data in real time is so voluminous that they need to be properly extracted to avoid bottlenecks and identify the desired traffic. A continuous flow of data is both a potential hurdle and an asset. While it may increase data traffic, the availability of an IoT/WSN device location depends on the amount of data flowing in real time. Many issues may arise when integrating in the WSN environment and can only be overcome if all possible limitations and complications are explored and assessed. Organizations must address these questions extensively before integrating WSNs with the Cloud Platform.

A number of human and physical resources are typically required to accomplish the operation and maintenance of a successful WSN-SN integration solution. In the scenario of a fully employable Cloud alternative, the organization must have enough dedicated resources to make acceptable decisions on the distribution of duties for cloud-based WSN. Using the dedicated services of trained personnel to establish and maintain a fully workable network has a high financial cost. The proper allocation of resources will drive many business leaders to decide to invest in the development of customized solutions that meet the requirements. The electronic system in which data is easily accessed by any person is secure and can be compromised to a variety of extents. Device and software manufacturers go to great lengths to protect customer privacy and personally identifiable information. While the integration is possible, it is essential that safety protocols be in place to protect valuable company assets from service interruption or data theft. Hostile advancement of network design can create havoc if it affects the operational practices WSNs are making.

## 5. Case Studies and Use Cases

This section includes case studies and use cases from various organizations that utilize the ServiceNow Cloud platform in integration with IEEE 802.15.4 network-based Wireless Sensor Networks (WSNs) in order to solve their specific security challenges. The interpretation and insight provided by each of the following organizations shed light upon the feasibility of our approach in managing the security of WSNs. Each use case entry is structured as follows, where possible, including the severity of the challenge and the solution provided through the adoption of the ServiceNow/WSN integration.

United Nations Office of Project Services (UNOPS) Challenge

The UNOPS wanted to monitor and secure the health and status of data centers reaching operational areas in countries where cable management and other data center requirements are negligible. The offices did not have sufficient infrastructure support [21]. Solution The UNOPS wanted an affordable, low-power solution that is non-intrusive and alerts the office in the event of overheating and hardware failure of the data center equipment. Today they use more than 130 sensors deployed in Iraq and Syria as their SaaS. The location of the sensors is treated with discretion for security reasons. The Regional IT Staff are currently using tools to analyze the results of the individual sensors and maintain the databases. This provides them with an automated response to any alarms. A typical hardware response would include a fan or turning the power supply off.

Solvera Information Challenge

To reduce the company's data leakage and to enhance the existing network security systems and procedures. Early warning and monitoring to enhance the protection of critical assets. Solution

The ServiceNow integration with the middleware allowed services to be written to take alerts from the WSN and pass these as events, with a unique reference number into ServiceNow's SaaS "Service Desk" and Server "Express," allowing us to build "Business Services," "Portfolio," and "Knowledge Base" structure and integrate these alerts directly into our business "Intelligent Operation Center" (IOC), where our security and network operation center analysts can action them in tier 1, 2, or on a flow-through basis.

# 6. Future Research Directions

The research we have previously conducted imparts several suggestions and cues for future research directions. Wireless technologies continue to expand, and new areas could be integrated with instead of, or along with, WSNs, specifically other sensor networks and other wireless networks such as IoT. Future research could include different attack scenarios and investigations on how these attacks could be prevented or mitigated by integrating the respective sensor networks with the cloud-based platform. Our research primarily used traditional security techniques, which include cryptography; however, other novel techniques such as the use of blockchain or quantum-resistant algorithms are emerging as the future direction of wireless security technologies. A new direction in research could be integrating the Security Operations Center with the cloud platform, as the future could present an environment where processes, data, and even technologies sit in the cloud.

Substantial inquiry is essential to propose an elaborated and well-established BLE-WiFi integration for WSNs. WSN is a pervasive system and may receive attacks in more advanced and innovative formats that are exclusive to WSN and IoT. Studies may focus on the novel modulated signals and their effects in WSN and require the development of a new mechanism to prevent the attack. Among the systems that can be used to integrate WSNs, was proposed. Studies may propose the use of other systems and demonstrate their impact on the integration of the security of WSN. These solutions may be compared via the use of suitable hygiene factors in the network and its integration and compared for positive and negative impacts. The efficiency of different security mechanisms, including lightweight security protocols, is a part of WSN technologies. The integration of these protocols can be investigated and explored, and the emphasis can be to develop an understanding of these systems. The continued investment in these frameworks is of paramount importance. An exploration and investigation of machine-learning methodologies could provide better solutions, especially in networks where sensor data contain results of sensory data in the form of low-variance sensors as previously utilized. The development of security mechanisms needs a multidisciplinary approach. The explicit use of physical security experts should be carried out when developing security protocols for WSNs. Likewise, a better understanding of the combined systems by including IoT systems may present a holistic security solution to the overall security of the integrated WSN/IoT systems. The characteristics of IoT systems, particularly the introductory section, are paramount, and it would be useful to understand the characteristics of these systems as part of future research, which could contribute to our current study and identify the ideal technologies to protect their system in the enhancement of our WSN systems that utilize in the cloud platforms.

# 7. Conclusion and Recommendations

The convergence of technological advancement in wireless communication technology using WSNs and central management has brought in the futuristic vision to strengthen the integration of WSNs with the versatile cloud platform. Collaboration with the cloud platform adds value to the integration where the cloud takes care of back-end infrastructure, security, server support, databases, hardware, and operating systems. The principal advantages of utilizing the platform are improved security, quality-of-service, operational efficiency, enhanced resource optimization, and protection from investment in-house expertise to accomplish various activities around the clock. In this research, we focused primarily on the security considerations of WSNs and the security details of the platform. Several recommendations have been made for integrating sensitive WSNs with the cloud platform. The suggestions include the necessity of careful and specific assessment of the liability and endpoints required for a trustworthy implementation in the long run; considering the business and service impact for clearances; the renovation of Access Management and Firewalls; the need to reconsider the capability and management of alerts and notifications; as well as a clear comprehension of the required Incident and Patch Management strategy. In conclusion, the need for successful completion of this WSN service integration is paramount in order to ensure system resilience. The future work of integrating WSNs with the system should categorize the kind of WSN platform in relation to systems that can incorporate tactic interfaces. To facilitate the long-term integration and implementation of the proposed WSN service, the subcommands such as a detailed filtration of the data packet, data inspection for intrusion, sharing of the event logs among WSN service, assessment of identifying flaws, as well as reconfiguring should be summarized with an examination of the proportion of execution time and the requirements for more modifications to these module components. In this new technological ecosystem, network security is becoming even more crucial and a tough fight and is striving to be flexible enough to deal with the especially steep hike in threat analysis possibilities. The topology of the WSNs is gigantic with several interconnected nodes. As a result of these volumes of nodes, the WSN can produce millions of alerts in a short span of time. Some of the contextual data can be unusual, thus not being a breach in any scenario. If mentioned a list of authorized exceptions, the WSN should comply against the list. Future research will be focused on identifying the broad classification manner for data packets based on edge and user traffic, and a thorough examination of the adaptive distribution and re-preservation of the cloud's resources.